\documentclass[prb,twocolumn,showpacs]{revtex4}
\usepackage{graphicx,graphics,psfrag,amsmath,calc,amssymb}

\begin{document}
\title{Vortex quasi-crystals in mesoscopic superconducting samples}
\author{Wei Zhang}
\altaffiliation{Current address: Department of Physics, University of Michigan, Ann Arbor, Michigan, 48109.}
\author{C. A. R. S\'{a} de Melo}
\affiliation{School of Physics, Georgia Institute of Technology,
Atlanta, Georgia 30332}

\date{\today}

\begin{abstract}

There seems to be a one to one correspondence between the phases of atomic
and molecular matter (AMOM) and vortex matter (VM)
in superfluids and superconductors.
Crystals, liquids and glasses have been experimentally
observed in both AMOM and VM.
Here, we propose a vortex quasi-crystal state which
can be stabilized due to boundary and surface energy effects
for samples of special shapes and sizes.
For finite sized pentagonal samples, it is proposed that a phase transition
between a vortex crystal and a vortex quasi-crystal occurs
as a function of magnetic field and temperature as the sample size
is reduced.

\end{abstract}
\pacs{74.25.Op, 74.25.-q, 74.25.Dw}

\maketitle

%
%% PACS number 74.25.Op = Mixed state, critical fields, and surface sheath
%% PACS number 74.25.-q = Properties of type I and type II superconductors
%% PACS number 74.25.Dw = Superconductivity phase diagrams
%

\section{Introduction}
\label{sec:introduction}

The general subject of vortex physics in superconductors is quite interesting
since there seems to be a large variety of possible equilibrium vortex
phases in superconductors~\cite{crabtree&nelson}.
The term ``vortex matter'' has been coined to emphasize
the complexity and diversity of vortex phases in superconductors
when compared to atomic and molecular matter.
One can think of a one to one correspondence between
phases in atomic and molecular matter (AMOM) and phases in vortex matter
(VM). A liquid in AMOM corresponds to a vortex liquid in VM~\cite{liquid-1,liquid-2};
a crystalline lattice in AMOM corresponds to a vortex lattice in VM~\cite{abrikosov};
an amorphous or glassy solid in AMOM corresponds to an amorphous
or glassy vortex system in VM~\cite{glass}.
In addition, a quasi-crystal is another very interesting state which
has been experimentally discovered in AMOM~\cite{experiments},
but there are no corresponding experiments for vortex matter.

The possibility of vortex quasi-crystals was initially discussed
several years ago~\cite{sademelo-99}, but more recently quasi-periodic
arrangements of vortices were discussed in the presence of underlying
quasiperiodic pinning potentials~\cite{misko-05, villegas-06, kemmler-06, silhanek-06}.
In these recent theoretical and experimental studies it was suggested that
the physical properties are quite unconventional due to the quasi-periodic pinning
potentials.  However, unlike these studies, the present paper is dedicated
to a discussion of vortex quasi-crystalline phases in superconductors without quasi-periodic pinning potentials.
This manuscript is an extension of our previous unpublished work~\cite{sademelo-99, wei-06},
where it was argued that vortex quasi-crystals maybe stabilized by boundary
effects and surface energies in finite systems. The present situation is
quite distinct from the case of quasi-periodic pinning of vortices~\cite{misko-05} and from the case
encountered in some atomic and molecular matter, where certain types of interactions can lead to
colloidal quasi-crystals~\cite{denton-98}.

The central question of this manuscript is: under what conditions
a quasi-crystalline arrangement of vortices is at all possible?
A definite possibility is to argue that a stable vortex quasi-crystal
can arise from an imposed quasi-periodic potential
like for instance in the case of a superconductor with
quasi-periodic pinning potentials~\cite{misko-05},
or in heterostructures consisting of a superconductor film
grown on top of a quasi-crystal film.
Another possibility is to create a quasi-periodic optical lattice
with laser beams, trap and cool atoms, and produce vortex
quasi-crystals in Bose or Fermi superfluids.
This experiment is natural since the production of vortex lattices in superfluid Bose or Fermi
ultra-cold atoms has become standard~\cite{cornell-99, ketterle-05}, and more
recently square optical lattices were used to induce transitions between triangular
and square vortex lattices using a mask technique~\cite{cornell-06}.
Thus, the generation of five-fold quasi-periodic optical
lattices using a mask with five holes as the vertices of a regular pentagon
(or other methods) may also be used to produce transitions
between triangular and five-fold symmetric vortex lattices.

However, in this manuscript, we concentrate on the possibility of
stabilizing vortex quasi-crystals only due to boundary effects in
finite superconducting samples, where the sample size and shape
play an important role. This option is motivated by recent
experimental studies of the vortex structure in disk,
triangular, square, and star-shaped mesoscopic
samples~\cite{geim-97,geim-00,chibotaru-01,dikin-03, berdiyorov-06}.
These works provide us with the technology that allows the preparation of
pentagonal (pentagon-cylinders) or decagonal (decagonal-cylinders)
samples as potential candidates to produce 5-fold vortex quasi-crystals.
Furthermore, in these mesoscopic systems of cross-sectional area $A$,
the number of vortices $N$ is essentially given by $N = H A/\Phi_0$,
where $\Phi_0$ is the quantum of flux, and $H$ is the magnetic field.
Therefore samples with large upper critical field $H_{c_2}$ may
produce a large enough number of vortices even the cross-sectional area is small,
and the appearance of quasi-crystalline order is a definite possibility.
For example, for a sample of cross sectional area $A = 1 {\rm \mu m}^2$ and upper critical
field at zero temperature $H_{c_2} (0) = 10{\rm T}$,
the number of vortices $N \approx 5000$ is quite substantial
close to $H_{c_2} (0)$. In addition, since superconductors
with large $H_{c_2}$ are usually associated with
short coherence lengths, the vortex quasi-crystalline phase proposed
is more likely to be observable in proper samples of
short coherence length superconductors at high magnetic fields and low temperatures.

This possibility is considered here under the following program.
In this manuscript only the case of an isotropic
type II superconductor in a magnetic field is considered. First,
the bulk free energy is calculated for a triangular, a square and a 5-fold
quasi-crystal array. The 5-fold quasi-crystal array is modeled
by a Penrose tiling of the plane as shown in Fig. 1.
It is shown that the Penrose tile array (vortex quasi-crystal) has a bulk free
energy which is just a few percent higher than the triangular array.
Second, instead of considering an infinite
(bulk) system, a pentagon cylinder sample is discussed. The pentagon cylinder
has a pentagonal cross-section (in the $xy$ plane) with side dimension
$\ell$, and with height $L$ along the ${\bf z}$ direction.
In this case, when the sample size gets smaller
the contribution of the boundaries (surface energy) to the total free energy
of the system becomes more important.
The surface energy is highly sensitive to the symmetry and to the surface
area of the boundaries. Taking into account the surface free energies,
it is shown that the Penrose tiling (vortex quasi-crystal) has lower
free energy than the triangular lattice in certain regions of
the magnetic field versus temperature phase diagram. This is
suggestive that a ``first order phase transition'' may occur between the
triangular lattice and the Penrose tiling
(vortex quasi-crystal)~\cite{footnote2}.

The remainder of this manuscript is organized as follows.
In section~\ref{sec:ginzburg-landau}, we discuss a Ginzburg-Landau approach to obtain
the free-energy of the triangular, square and Penrose-tiling in the bulk.
In section~\ref{sec:boundaries}, we present the effects of pentagonal boundaries on
the free energy of the triangular lattice and Penrose-tiling, and obtain the critical magnetic
field where the transition from triangular vortex lattice to a vortex quasi-crystal occurs.
In section~\ref{sec:thermodynamics}, we discuss thermodynamic quantities like the change
in magnetization and entropy, while in section~\ref{sec:neutron-scattering}, we
discuss neutron scattering, Bitter decoration and scanning tunneling microscopy
as possible probes to identify the quasi-crystalline structure.
Lastly, in section~\ref{sec:conclusions} we state our conclusions and summarize.
\begin{figure}[ht]
\begin{center}
\includegraphics[width=7.0cm]{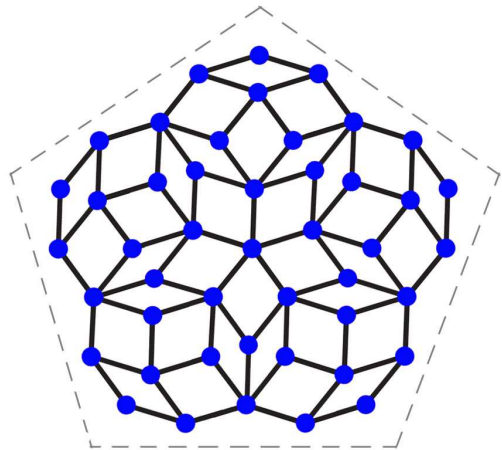}
\caption{We consider a vortex lattice in a pentagon superconducting sample
having Penrose tiling with 5-fold rotational symmetry.
The tile consists of two types of lozenges, one with internal
angles 36$^\circ$ and 144$^\circ$ (thin lozenge) and
the other with internal angles 72$^\circ$ and 108$^\circ$ (thick lozenge).}
\label{penrose}
\end{center}
\end{figure}

\section{Ginzburg-Landau theory}
\label{sec:ginzburg-landau}

The starting point of our analysis to describe the vortex-quasicrystal state
is the Ginzburg-Landau (GL) free energy density
\begin{equation}
\label{eqn:fed}
\Delta F_s = F_1
+ \frac{1}{2m} \left\vert \left( -i\hbar \nabla - \frac{2e {\bf A}}{c} \right)
\Psi ({\bf r}) \right\vert^2 + \frac{H^2}{8 \pi}
\end{equation}
for a bulk isotropic superconductor with no disorder,
where $\Delta F_s = F_s - F_n$ is the free energy density
difference between the superconducting state $(F_s)$ and the normal state $(F_n)$,
and $F_1 = \alpha \vert \Psi ({\bf r}) \vert^2 + \beta  \vert \Psi ({\bf r}) \vert^4/2$.
It is useful to introduce the dimensionless quantities
$f = \Psi \sqrt{-\beta/\alpha}$, $\mbox{\boldmath$\rho$} = {\bf r}/ \lambda (T)$,
${\cal A} = 2 \pi \xi (T) {\bf A}/ \Phi_0 $, and
${\cal H} = 2 \pi \xi (T) \lambda (T) H / \Phi_0$,
where $\lambda$ is the penetration depth, $\xi$ is the coherence length,
and $\Phi_0 = hc/2e$ is the unit flux.
Here, $f = f_0 \exp (i\phi)$, and ${\cal A } = {\cal A}_0 +  {\nabla \phi / \kappa}$ with
$\kappa = \lambda /\xi$ is the GL parameter.
Notice that when $f_0 = 1$ the system is fully superconducting,
and when $f_0 = 0$ the system is normal, thus $f_0 \le 1$ always.
Considering the minimization of free energy density with respect to
${\cal A}$ and $\Psi$ it is easy to arrive at equations for
the dimensionless functions ${\bf {\cal H}}$ and $f_0$.
The microscopic field is
\begin{equation}
\label{eqn:hsolution}
{\cal H}(x,y) = \kappa \left[ 1 + \frac{ H - H_{c_2}}{H_{c_2} } \right]
- \frac{g(x,y)}{ 2 \kappa},
\end{equation}
with ${\bf H}$ parallel to the ${\bf z}$ direction,
and the equation for $f_0$ can be written as
\begin{equation}
\label{eqn:glinear}
\nabla^2 (\log g) + 2\kappa^2 = 0,
\end{equation}
where $g = f_0^2$ is a positive definite function.
Notice that ${\cal H} = \kappa$ for $H = H_{c_2}$ and $f_0 = 0$ $(g = 0)$.
The most general solution of Eq.~(\ref{eqn:glinear}) has the form~\cite{saint-james-69}
\begin{equation}
\label{eqn:gsolution}
g(x,y) = \exp\left[-\kappa^2(x^2 + y^2)/2\right] \exp \left[ \gamma(x,y) \right],
\end{equation}
where $\gamma(x,y)$ satisfies Laplace's equation $\nabla^2 \gamma (x,y) = 0$.
This means that $\gamma (x,y)$ is a harmonic function
excluding the locations of vortices, and can be expressed
as the real part of an analytic function of $z = x + iy$.
This observation has very important consequences
for the microscopic field profile ${\cal H} (x, y)$ of Eq.~(\ref{eqn:hsolution}),
which depends strongly on the structure of $g(x,y)$.

The bulk Gibbs free energy density is given by~\cite{abrikosov}
\begin{equation}
\label{eqn:gibbs}
G_s (H,T) = G_n (H,T) - \frac{1}{8 \pi} \frac{ (H_{c_2} - H)^2} {(2 \kappa^2 - 1) \beta},
\end{equation}
where the parameter $\beta = {\langle g^2 \rangle/ {\langle g \rangle}^2}$
is a geometrical factor independent of $\kappa$.
The notation $\langle \cdots \rangle$ indicates average over volume.
It is important to notice that $\beta \ge 1$ no matter what is the
form of $g(x,y)$ because of the Schwartz inequality. In addition, notice
that the Gibbs free energy above is a minimum, whenever $\beta$ reaches its minimum value.

For the purpose of calculating the parameter $\beta$ and the free energies
corresponding to different vortex configurations,
the analytical structure of $g(x,y)$ in the complex plane is used to
rewrite it as
\begin{equation}
\label{eqn:gzzbar}
g(z,{\bar z}) = \exp (-\kappa^2 z {\bar z}/2) \vert P (z) \vert,
\end{equation}
where $P (z) = {\cal N} \prod_{i = 1}^M (z - z_i)$.
Here, each $z_i$ corresponds to a zero of $g$ in the complex plane,
and $M$ is the number of zeros. The zeros $z_i$ indicate the location
of vortices. From now on it is assumed that there is only one vortex
with flux $\Phi_0$ at each position $z_i$, i.e., each zero is non-degenerate.
In this case, $M$ corresponds to the number of vortices,
and thus the total flux threading the sample is $\Phi = M \Phi_0$.
The normalization coefficient ${\cal N}$ just guarantees that $g(z,{\bar z}) \le 1$.
Depending on the locations of the zeros of $g(z, \bar z)$,
it is possible to study several possibilities of periodic and quasi-periodic vortex arrangements.
In this study only vortex crystals corresponding to
triangular and square lattices and vortex quasi-crystals corresponding
to the 5-fold Penrose tiling of the plane are considered.
Both the square lattices and triangular lattices can be generated via
the tiling method, i.e., via the periodic arrangements of identical
square tiles or identical lozenges of internal angles ($60^\circ$ and $120^\circ$).
The Penrose lattice, however, requires quasi-periodic
arrangements of two types of tiles (lozenges), one with
internal angles $36^\circ$ and $144^\circ$ and the other with internal
angles $72^\circ$ and $108^\circ$. Using the representation in Eq. (\ref{eqn:gzzbar}),
the values of $\beta$ for the triangular, square and Penrose tiling are
respectively $\beta_3 = 1.16$, $\beta_4 = 1.18$ and $\beta_5 = 1.22$. This
immediately indicates that the triangular lattice has lower free energy
than the square lattice which has lower free energy
than the 5-fold vortex quasi-crystal (Penrose tiling), as expected.
However, the fact that the free energy difference
between the triangular and 5-fold vortex quasi-crystal is only a few percent
suggests that appropriate boundaries can favor 5-fold symmetry as the sample size
gets smaller, as can appropriately imposed quasi-periodic potentials.
Thus next, we discuss boundary effects and the magnetic field versus temperature
phase diagram.

\section{Boundary effects and phase diagram}
\label{sec:boundaries}

In order to investigate how boundary effects can modify the total free
energy of the system, a sample in the shape of a pentagon cylinder of
side $\ell$ and height $L$ is considered.
Imposing that no currents flow through the sample boundaries
leads to the condition
\begin{equation}
\label{eqn:normal}
{\bf \hat n \cdot} \left[
{ { - i {\bf \nabla}} - {2\pi {\bf A} / \Phi_0}}\right] \Psi(x,y) = 0
\end{equation}
at all five side faces. The unit vector ${\bf \hat n}$ points along the
normal direction of each facet of the pentagon cylinder.
The boundary conditions can be translated
in terms of the harmonic function $\gamma (x,y)$ as
\begin{equation}
\label{eqn:bound-gamma}
n_x \frac{\partial \gamma}{\partial y} -
n_y \frac{\partial \gamma}{\partial x}
= \kappa (n_x y - n_y x),
\end{equation}
where $n_x$ and $n_y$ are the $x$ and $y$ components of the normal unit
vector ${\bf \hat n}$ at each one of the pentagon cylinder side faces.
The solution for this boundary value problem can be obtained
using a Schwarz-Christoffel conformal map of the pentagon into a semi-infinite plane
\begin{equation}
\label{eqn:sc-transf}
\frac{dz}{d w} = C (w - x_1)^{-2/5} (w^2 - x^{2}_2)^{-2/5}
(w^2 - x^{2}_3)^{-2/5},
\end{equation}
where the vertices of the pentagon located at $z_{i}$ are mapped into the
points $(x_1, \pm x_2, \pm x_3)$ on the real axis of the $w$-plane.
The full solution of this problem is complicated,
and requires heavy use of numerical methods~\cite{henrici}.
However, the free energy density can be estimated when the bulk solution
in Eq. (\ref{eqn:gzzbar}) is treated as a variational solution of
the boundary value problem determined by Eqs. (\ref{eqn:glinear}),
(\ref{eqn:bound-gamma}) and (\ref{eqn:sc-transf}).

The Gibbs free energy density difference $\Delta G = G_3 - G_5$
between the triangular and the 5-fold quasi-periodic Penrose structure
then becomes
\begin{equation}
\label{eqn:gibbs-dif}
\Delta G = - \frac{1}{8\pi} \frac{(H_{c_2} - H)^2}{(2\kappa^2 - 1) \beta^*}
+ \frac{H_c^2}{4 \pi} \epsilon(H),
\end{equation}
where $\beta^* = \beta_3 \beta_5/(\beta_5 - \beta_3)$,
$H_c$ is the thermodynamic critical field, and
$\epsilon= (\alpha_3 - \alpha_5)(2 R_e + \alpha_3 + \alpha_5)/2 R_{e}^2$,
with $\alpha_3 = 0.93 a_0$, $\alpha_5 = 0.90 a_0$, and
$a_0  = \sqrt{\Phi_0/H} $.
The effective length of the sample $R_{e} = \ell/\sqrt{4 - \tau^2}$,
where $\tau = 2 \cos(\pi/5)$ is the golden mean.
Here, $R_e$ corresponds to the radius of the circle that
circumscribes the pentagonal sample.

This expression for $\Delta G$ is valid only when $H \gg \Phi_0/R_e^2$.
The second term in $\Delta G$ takes into account the boundary mismatch energy,
and indicates that as the size of the pentagon cylinder gets smaller
it becomes more favorable to have a 5-fold quasi-crystal rather than a regular
triangular lattice. Notice, however, that when $R_e \to \infty$ the
triangular lattice has lower Gibbs free energy as it must, and no transition
to a 5-fold quasi-crystal occurs. Thus, this possible transition may occur
for finite sized samples only. From the condition that $\Delta G = 0$ we obtain
\begin{equation}
\label{eqn:hq}
H_Q = H_{c_2}
\left[ 1 - \kappa^* \sqrt{ \beta^* \epsilon(H_Q)} \right],
\end{equation}
where the transition to a quasi-crystal occurs.
Here, $\kappa^* = \sqrt{2 \kappa^2 - 1}/\kappa$.
\begin{figure}[ht]
\begin{center}
\includegraphics[width=7.0cm]{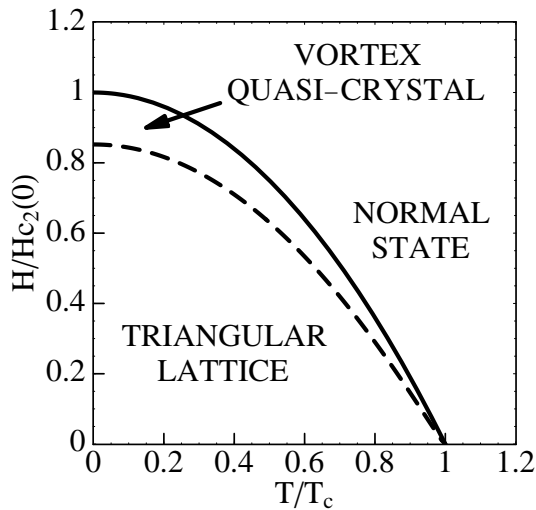}
\caption{$H - T$ phase diagram for a pentagonal cylinder cut out of
a regular cylinder of radius for $R_e = 10^{-6}$m.
The solid and dashed lines represent $H_{c_2}$ and $H_Q$, respectively,
while $H_{c_1}$ is not shown.
The superconductor is assumed to have $\kappa = 20$, $H_{c_2} (0) = 10$T.
}
\label{fig:1}
\end{center}
\end{figure}

\begin{figure}[ht]
\begin{center}
\includegraphics[width=7.0cm]{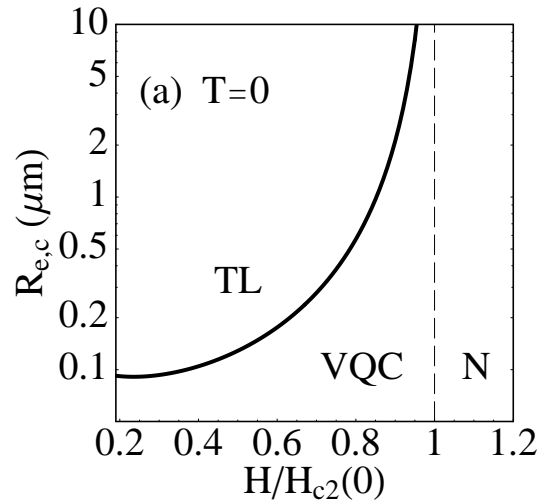}
\includegraphics[width=7.0cm]{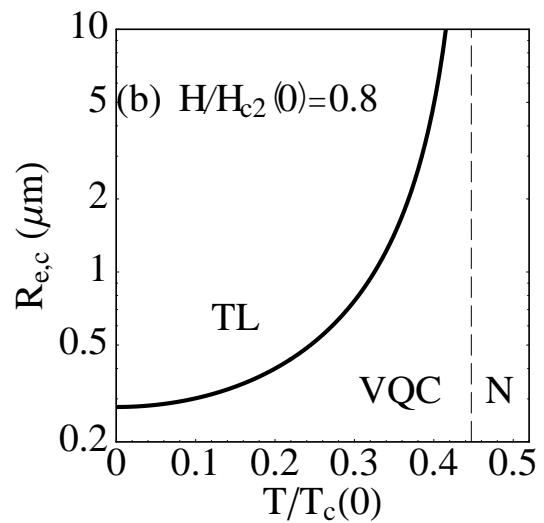}
\caption{Critical value of superconducting sample size $R_{e,c}$ as
a function of (a) $H$ for a given $T<T_c(0)$,
and as a function of (b) $T$ for a given $H<H_{c_2}(0)$.
Parameters used in these plots are the same as in Fig.~\ref{fig:1},
and the regions labeled TL, VQC, and N correspond to triangular lattice,
vortex quasi-crystal, and normal phases, respectively.}
\label{fig:Rec}
\end{center}
\end{figure}

The phase diagram for a superconductor with $\kappa = 20$,
$H_{c_2} (0) = 10$ T and $R_e = 10^{-6}$ m is shown~\cite{hc3} in Fig.~\ref{fig:1}
using $H_{c_2} (T) = H_{c_2} (0) \left[ 1 - (T/T_c)^2 \right]$.
Notice that the number of vortices $N$ close to $H_{Q}$ can be quite large
at low temperatures. For the parameters above $N = H_Q A/\Phi_0 \approx 4000$,
and the vortex quasi-crystal structure can be extracted.
However, the phase diagram becomes less precise at high temperatures,
since the low value of $H$ sets a restriction on the number of vortices.
Furthermore, for fixed $H_{c_2} (0)$, one can find the critical value of $R_{e,c}$
below which the vortex quasi-crystal phase appears. Curves of $R_{e,c}$ versus $T$
or $H$ can be obtained and are shown in Fig. \ref{fig:Rec}.
Notice that the critical $R_{e,c}$ increases with increasing $H$ ($T$) for a fixed $T$ ($H$).
Notice also that for a fixed sample size, and fixed magnetic field the vortex-quasi crystal
phase always occurs at higher temperatures due to its higher entropy.

In order to address the signature of the crystal-quasicrystal
phase transition, we discuss next thermodynamic properties
including magnetization and entropy of the triangular
and Penrose lattices in the pentagonal geometry.

\section{Thermodynamics: changes in magnetization and entropy}
\label{sec:thermodynamics}

\begin{figure}[ht]
\begin{center}
\hspace{-1cm}
\includegraphics[width=7.0cm]{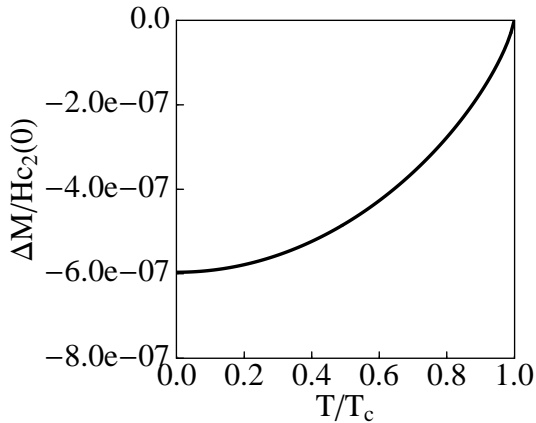}
\caption{The jump discontinuity $\Delta M = M_3 - M_5$ at
the critical field $H_Q$ for various reduced temperatures $T/T_c$,
using the same parameters of Fig. \ref{fig:1}.}
\label{fig:2}
\end{center}
\end{figure}

The jump of the magnetization $\Delta M (= M_3 - M_5)$
as a function of temperature at the critical field $H_Q$ can be calculated
from the Gibbs free energy leading to
\begin{equation}
\label{eqn:mag35}
\Delta M = \frac{H_{c_2}}{4\pi} \left [ -\frac{\gamma^*}{(2\kappa^2 -1)\beta^*}
+
\Delta M_s \right],
\end{equation}
where $\gamma^* = \kappa^* \sqrt{\beta^* \epsilon}$ and
$\Delta M_s = (\alpha_3 - \alpha_5)(R_e + \alpha_3 + \alpha_5)/4\kappa^2 R_e^2 (1-\gamma^*)$.
A plot of $\Delta M$ is illustrated in Fig. \ref{fig:2} for
the same parameters of Fig. \ref{fig:1}.
Using these parameters produces jump discontinuities
$\Delta M \approx - 0.060 $G at $T = 0$, and
$\Delta M \approx - 0.028 $G at $T = 0.8 T_c$.
However, measurements of $\Delta M$ may require the preparation of an
array of identical pentagonal cylinders to enhance the overall value.
Notice that $\Delta M < 0$ indicates that the 5-fold vortex quasi-crystal
is denser than the triangular vortex lattice at $H_Q$,
being at best a few percent denser at $T = 0$.

In addition to magnetization measurements, it is also interesting
to perform calorimetric experiments. However, specific heat measurements
are very difficult because they require large samples.
Since sample size is important for the present discussion it
is not clear that such experiments can be successfully performed.
Nevertheless, the thermodynamic relationship between the magnetization and
entropy jumps is revealed in the Clapeyron equation
\begin{equation}
\label{eqn:entropy}
\Delta S = S_3 - S_5 = - \Delta M dH_Q/dT.
\end{equation}
Since $d H_Q/ dT < 0$, and $\Delta M < 0$ implies that $\Delta S < 0$,
the entropy $S_3$ of the triangular vortex
lattice is less than the entropy $S_5$ of the 5-fold vortex quasi-crystal,
indicating that latent heat $L = T \Delta S$ is required to cause this phase transition.

Thermodynamic quantities can provide a good understanding of properties
averaged over the entire sample, and can characterize the signature of
the crystal-quasicrystal phase transition. However, as discussed above,
a good measurement of these quantities may require an array of identical samples,
which introduces experimental complexity and difficulty.
As another possibility, the use of local probes discussed
next is much desired in order to reveal the change in structure from
a triangular vortex crystal to a 5-fold vortex quasi-crystal.
For instance, neutron scattering, Bitter decoration or
scanning tunneling microscopy (STM) experiments may help
elucidate the structure of the vortex arrangement in mesoscopic samples.

\section{Local Probes: neutron diffraction and scanning tunneling microscopy}
\label{sec:neutron-scattering}

In neutron diffraction experiments periodic or
quasi-periodic variations of ${\cal H} (x, y)$ will result in Bragg peaks.
The position of these peaks determine the characteristic length scale
of the vortex structure and its symmetry. The neutron scattering amplitude in
the Born approximation is
\begin{equation}
\label{eqn:scattering}
b ({\bf q}) = \frac{M_n} {2\pi \hbar^2}
\int \mu_n H ({\bf r}) \exp(i {\bf q \cdot r} ) d{\bf r},
\end{equation}
where $\mu_n = 1.91 e \hbar/M_n c$ is the neutron magnetic moment and the
$M_n$ is the neutron mass. The scattering amplitude $b ({\bf q})$ is directly
proportional to the Fourier transform $H ({\bf q})$ [${\cal H} (q_x, q_y)$]
of the microscopic field $H ({\bf r})$ [${\cal H} (x,y)$] of Eq. (\ref{eqn:hsolution}).
The neutron scattering cross section
\begin{equation}
\label{eqn:cross-section}
\sigma (q_x, q_y) = 4\pi^2 \vert b (q_x, q_y) \vert^2
\end{equation}
has sharp peaks at $ (q_x, q_y) = (0,0) $ (central peak)
and at $(q_x, q_y) = (\pm q_{xNm}, \pm q_{yNm}) $ (first Bragg peaks),
where $q_{xNm} = Q_N \cos (m\pi/N)$ and $q_{yNm} = Q_N \sin (m\pi/N)$,
with $ m = 0, 1, ..., N - 1$. For the triangular lattice
$N = 3$ the first Bragg peak occurs at
$|Q_3| = 2.31 \times \pi/d_3$, where $d_3$ is the lattice spacing.
For the 5-fold vortex quasi-crystal
(Penrose Lattice, $N = 5$) the first Bragg peak occurs at
$|Q_5|= 2.46 \times \pi/ d_5$, where $d_5$ is the side of a tile.
Since the sample size is important for the observation of a 5-fold
quasi-crystal, neutron scattering experiments may be difficult to
perform.

\begin{figure}[ht]
\begin{center}
\includegraphics[width=3.0cm]{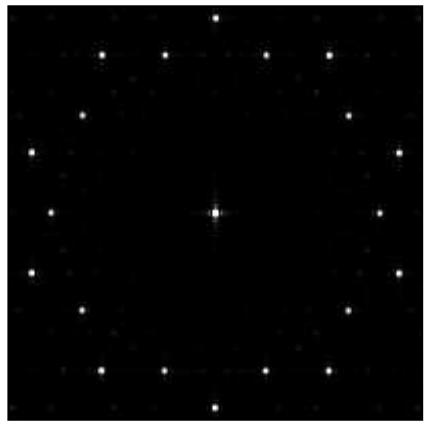}
\caption{First peaks of the square of the Fourier transformed (FT) pattern for the 5-fold vortex
quasi-crystal (Penrose lattice). Notice that the pattern is 10-fold symmetric, unlike
the case of the triangular vortex lattice, where the pattern is 6-fold symmetric.}
\label{fivefold}
\end{center}
\end{figure}

However, Bitter decoration might be a useful technique
if magnetic nanoparticles could be used to decorate the magnetic
field profile, and then be seen by a scanning tunneling microscope (STM)
(magnetic or non-magnetic). Furthermore, it may be possible to use
just an STM to scan over the pentagonal sample and probe the local
density of states which is substantially different inside and outside of vortex
cores, due to the presence of vortex cores states. In this case, it may
be also useful to make a periodic pattern of pentagonal samples,
to obtain an ensemble average.
It should be possible as well
to perfom STM scans at different fields and temperatures
in the vicinity of $H_Q (T)$, which
would reveal the real space locations of vortices.
The pattern obtained could then be Fourier transformed (FT) to obtain a 6-fold pattern
for the triangular vortex lattice and a 10-fold pattern for
the 5-fold vortex quasi-crystal (Penrose lattice), which is shown
in Fig.~\ref{fivefold}.

Now that the phase diagram, thermodynamics, and the local signatures
of a vortex quasi-crystal have been discussed, it is important to say a word on
the stability of such structures. A stability analysis in the free
energy can be performed by moving the vortices away from their
equilibrium positions $z_i$ to $z_i + \delta z_i$.
The eigenvalues associated with these displacements indicates that
for $H > H_Q (T)$ the vortex quasi-crystal lattice is stable,
and therefore potentially realizable in mesoscopic samples.

\section{Final comments and conclusions}
\label{sec:conclusions}

Before concluding, we would like to make several comments in connection
with the approximations used and the observability of the vortex quasi-crystal
phase discussed.

First, it should be emphasized that our free energy analysis
provides a preliminary understanding of the vortex quasi-crystal phase,
but further detailed numerical work is necessary.
For instance, the vortex quasi-crystal sitting on a Penrose lattice
is only one possible state in a pentagon superconducting sample.
Although we have shown that this state has lower free energy
than a triangular or rectangular lattice, we cannot rule out
other possibilities based on the present calculation.

Second, since there is no full compatibility of the 5-fold
Penrose lattice with the pentagon cylinder geometry, the appearance
of disclinations and dislocations is possible.
As a result there is an additional possibility of a solid
(crystal or quasi-crystal) vortex structure in the center of
the sample, which melts or gets disordered at the sample boundaries~\cite{cabral-04}.

Third, the number of vortices in superconducting samples is restricted
by $N_{\rm max} = H_{c_2} (0) A/ \Phi_0$, which means that
for superconductors with high $H_{c_2}$ a large number of
vortices is possible. One can make a simple estimate for $N_{\rm max}$
by using the Ginzburg-Landau relation $H_{c_2} = \Phi_0/( 2\pi \xi^2 )$,
leading to $N_{\rm max} \sim [R_e/\xi(T=0)]^2$.
For conventional type II superconductors, the number $N_{\rm max}$
is only about $10^1 \sim 10^2$, thus the thermodynamic limit and
hence the definition of the quasi-crystal phase is questionable.
However, for materials with short coherence length (and large upper critical fields),
such as high-$T_c$ superconductors, where $\xi (T=0)$ is about $10^1 \sim 10^2$\AA,
then $N_{\rm max}$ can be as large as $10^4 \sim 10^5$,
and the quasi-crystalline structure can be well defined and observed.
Thus, the observation of a vortex quasi-crystal state is more likely to occur in
short coherence length superconductors at low temperatures and high
magnetic fields.

Finally, strong disorder in the sample can also destroy the quasi-crystal
structure due to the pinning of vortices, however, clean
mesoscopic superconducting materials already exist and shell structures
have been observed for Niobium samples of $\mu m$ sizes~\cite{grigorieva-06}.
Thus, we suspect that experimentally this should not be an issue
as vortex lattices (triangular) are routinely observed in reasonably clean superconducting samples.
Therefore, the choice of pentagonal mesoscopic samples of superconductors
with sufficiently large $H_{c_2}$ should allow for the observation of
the vortex quasi-crystal state.

In summary, we have shown that vortex quasi-crystals may be experimentally
observed in mesoscopic samples of type II superconductors with large upper
critical fields (short coherence lengths).
By taking into account boundary effects, sample shape and size,
we proposed that a first order phase transition occurs between a vortex
crystal and a vortex quasicrystal, as magnetic field and
temperature are varied.

We would like to thank Wai Kwok and Franco Nori for references,
the Aspen Center for Physics for their hospitality,
and NSF (DMR-0304380) for financial support.


\begin{references}
\bibitem{footnote2} The term ``first order phase transition'' appear in
quotes since the sample size is finite.
\bibitem{hc3} The critical field $H_{c_3} (T)$ for surface superconductivity
is not shown in Fig.~\ref{fig:1}.
\end{references}
\end{document}